# An accurate numerical method for systems of differentio-integral equations associated with multiphase flow

Jian-Jun SHU and Graham WILKS

School of Mechanical & Aerospace Engineering, Nanyang Technological University,
50 Nanyang Avenue, Singapore 639798
mjjshu@ntu.edu.sg
http://www.ntu.edu.sg/home/mjjshu

**Abstract.** A very simple and accurate numerical method which is applicable to systems of differentio-integral equations with quite general boundary conditions has been devised. Although the basic idea of this method stems from the Keller Box method, it solves the problem of systems of differential equations involving integral operators not previously considered by the Keller Box method. Two main preparatory stages are required: (i) a merging procedure for differential equations and conditions without integral operators and; (ii) a reduction procedure for differential equations and conditions with integral operators. The differencing processes are effectively simplified by means of the unit-step function. The nonlinear difference equations are solved by Newton's method using an efficient block arrow-like matrix factorization technique. As an example of the application of this method, the systems of equations for combined gravity body force and forced convection in laminar film condensation can be solved for prescribed values of physical constants.

## 1 Introduction

There are a large variety of numerical methods which are used to solve mathematical physical problems. Two particular methods, the Box scheme and the Crank-Nicolson scheme, seem to dominate in most practical applications. Keller [1] himself preferred and stressed the Box scheme. This scheme was devised in Keller [2] for solving diffusion problems, but it has subsequently been applied to a broad class of problems. It has been tested extensively on laminar flows, turbulent flows, separating flows and many other such problems [3]. These have almost invariably been 'one-boundary-layer flow' problem which are described only by differential equations and differential boundary conditions. The numerical method has not previously been adapted to multilayer problem which are more naturally described as differentio-integral equations and conditions.

In engineering, there are many multilayer problems involving the presence of distinct phases or immiscible fluids. The 'guess' method has usually been used to solve the problem [4]. The general procedure by which solutions are obtained may be outlined as follows: The values of some parameters are guessed so that the solutions can be carried out from one layer to another using a standard numerical method for one-layer flows. A check is then made as to whether the guessed values are correct. If not, a new guess is made and the entire calculation is repeated until the acceptable, correct values are obtained. The chief drawback of this method is its basis in guesswork. A particular problem may be sensitive to the correct guess and, moreover, it is difficult to adopt this method to non-similar flows.

For non-similar multilayer flows, systems of differentio-integral equations often need to be solved. Here a very simple and accurate numerical scheme which is applicable to quite general single or multiple boundary layer flow problems has been devised. The method shall be illustrated by showing its application in some detail to nonsimilar plane laminar incompressible double boundary layers and in particular to the equations for combined gravity body force and forced convection in laminar film condensation [5]. There is no difficulty in adopting the methods to laminar flows involving more than two layer flows and to turbulent flows (using the eddy viscosity and eddy conductivity formulations), but such calculations are not included here.

For incompressible laminar double boundary layer flow over a plane surface, the double-boundary-layer equations can be reduced to the dimensionless forms [5]:

$$\frac{\partial^3 f}{\partial \eta^3} + \alpha(\xi) f \frac{\partial^2 f}{\partial \eta^2} + \beta(\xi) \left[ \gamma(\xi) - \left( \frac{\partial f}{\partial \eta} \right)^2 \right] = 2\xi p(\xi) \left[ \frac{\partial f}{\partial \eta} \frac{\partial^2 f}{\partial \xi \partial \eta} - \frac{\partial^2 f}{\partial \eta^2} \frac{\partial f}{\partial \xi} \right]$$

$$\frac{1}{P_r} \frac{\partial^2 \theta}{\partial \eta^2} + \alpha(\xi) f \frac{\partial \theta}{\partial \eta} = 2\xi p(\xi) \left[ \frac{\partial f}{\partial \eta} \frac{\partial \theta}{\partial \xi} - \frac{\partial \theta}{\partial \eta} \frac{\partial f}{\partial \xi} \right] \qquad 0 < \eta < \eta_\delta(\xi)$$

$$\frac{\partial^3 f^*}{\partial \eta^{*3}} + \alpha(\xi) f^* \frac{\partial^2 f^*}{\partial \eta^{*2}} - \beta(\xi) \left( \frac{\partial f^*}{\partial \eta^*} \right)^2 = 2\xi p(\xi) \left[ \frac{\partial f^*}{\partial \eta^*} \frac{\partial^2 f^*}{\partial \xi \partial \eta^*} - \frac{\partial^2 f^*}{\partial \eta^{*2}} \frac{\partial f^*}{\partial \xi} \right] \qquad \eta^* > 0$$

$$H_0 \left[ \left( \frac{\partial \theta}{\partial \eta} \right)_{\eta=0} + P_r \alpha(\xi) \int_0^{\eta_\delta(\xi)} \frac{\partial f}{\partial \eta} \theta \, d\eta + 2\xi P_r p(\xi) \int_0^{\eta_\delta(\xi)} \left( \frac{\partial^2 f}{\partial \eta \partial \xi} \theta + \frac{\partial f}{\partial \eta} \frac{\partial \theta}{\partial \xi} \right) d\eta \right]$$
$$+ \alpha(\xi) (f)_{\eta=\eta_\delta(\xi)} + 2\xi p(\xi) \left( \frac{\partial f}{\partial \xi} \right)_{\eta=\eta_\delta(\xi)} + 2\xi p(\xi) \left( \frac{\partial f}{\partial \eta} \right)_{\eta=\eta_\delta(\xi)} \frac{d\eta_\delta(\xi)}{d\xi} = 0$$

where $\xi \geq 0$ is a transformed streamwise variable. $\eta$, $0 < \eta < \eta_\delta(\xi)$, measures distance in the inner boundary layer whose thickness is $\eta_\delta(\xi)$, an unknown function of $\xi$. $\eta^* \geq 0$ measures distance in the outer boundary layer. $f(\xi, \eta)$ and $f^*(\xi, \eta^*)$ are proportional to the inner and the outer stream functions respectively. $\theta(\xi, \eta)$ is dimensionless temperature in the inner boundary layer. $\alpha(\xi)$, $\beta(\xi)$, $\gamma(\xi)$ and $p(\xi)$ are the physical parameters, which depend on $\xi$. $H_0$ is a physical constant and $P_r$ denotes the Prandtl number.

The most general boundary conditions are of the forms

$$f(\xi,0) = 0, \quad \frac{\partial f(\xi,0)}{\partial \eta} = 0, \quad \theta(\xi,0) = 1, \quad \theta(\xi,\eta_\delta(\xi)) = 0,$$

$$C_0 f^*(\xi,0) = f(\xi,\eta_\delta(\xi)), \quad C_1 \frac{\partial f^*(\xi,0)}{\partial \eta^*} = \frac{\partial f(\xi,\eta_\delta(\xi))}{\partial \eta},$$

$$C_2 \frac{\partial^2 f^*(\xi,0)}{\partial \eta^{*2}} = \frac{\partial^2 f(\xi,\eta_\delta(\xi))}{\partial \eta^2}, \quad \frac{\partial f^*(\xi,+\infty)}{\partial \eta^*} = U_e(\xi)$$

where $C_0$, $C_1$ and $C_2$ are the physical constants and $U_e(\xi)$ is the external velocity field. The last boundary conditions, which provide the prerequisites for the solvability of the equations, are

$$\frac{\partial^2 f^*(\xi,+\infty)}{\partial \eta^{*2}} = 0 \quad \text{and} \quad \frac{\partial^3 f^*(\xi,+\infty)}{\partial \eta^{*3}} = 0.$$

Using these, it is easily shown from the equations that

$$-\beta(\xi)U_e(\xi) = 2\xi p(\xi)\frac{dU_e(\xi)}{d\xi}. \tag{1}$$

Therefore $U_e(\xi)$ is a solution of the above first order equation with at most an arbitrary constant of integration. This constant may be determined by the value of $U_e(0)$. When $\xi > 0$, $U_e(\xi)$ is a function determined by both the coefficients of the equations and the value of $U_e(0)$ rather than an arbitrary function. For this reason, the last boundary condition may be rewritten as

$$\frac{\partial f^*(0,+\infty)}{\partial \eta^*} = U_e(0) \quad \text{and} \quad \frac{\partial^2 f^*(\xi,+\infty)}{\partial \eta^{*2}} = 0 \quad \xi > 0.$$

Notice that the first, second and third of the governing equations are differential equations, whilst the fourth equation is a differentio-integral equation. The differencing scheme shall be separately examined for the purely differential systems before examining a scheme for the differentio-integral system.

## 2 Differential Merging Scheme

In order to make all unknown functions appear as explicit forms in both equations and conditions, the new variables are defined

$$F(\xi,\phi) = f(\xi,\eta), \quad \Theta(\xi,\phi) = \theta(\xi,\eta), \quad \phi = \eta/\eta_\delta(\xi),$$

$$F^*(\xi,\phi^*) = f^*(\xi,\eta^*), \quad \phi^* = 1 + \eta^*/\eta_\delta(\xi).$$

J.-J. Shu and G. WilksThe differential equations and all boundary conditions transform to the following systems of equations

$$\frac{\partial^3 F}{\partial \phi^3} + \alpha(\xi)\eta_\delta(\xi)F\frac{\partial^2 F}{\partial \phi^2} + \beta(\xi)\eta_\delta(\xi)\left[\gamma(\xi)\eta_\delta^2(\xi) - \left(\frac{\partial F}{\partial \phi}\right)^2\right]$$

$$= 2\xi p(\xi)\left[\eta_\delta(\xi)\frac{\partial F}{\partial \phi}\frac{\partial^2 F}{\partial \xi \partial \phi} - \frac{d\eta_\delta(\xi)}{d\xi}\left(\frac{\partial F}{\partial \phi}\right)^2 - \eta_\delta(\xi)\frac{\partial^2 F}{\partial \phi^2}\frac{\partial F}{\partial \xi}\right]$$

$$\frac{1}{P_r}\frac{\partial^2 \Theta}{\partial \phi^2} + \alpha(\xi)\eta_\delta(\xi)F\frac{\partial \Theta}{\partial \phi} = 2\xi p(\xi)\eta_\delta(\xi)\left[\frac{\partial F}{\partial \phi}\frac{\partial \Theta}{\partial \xi} - \frac{\partial \Theta}{\partial \phi}\frac{\partial F}{\partial \xi}\right] \quad 0 < \phi < 1$$

$$\frac{\partial^3 F^*}{\partial \phi^{*3}} + \alpha(\xi)\eta_\delta(\xi)F^*\frac{\partial^2 F^*}{\partial \phi^{*2}} - \beta(\xi)\eta_\delta(\xi)\left(\frac{\partial F^*}{\partial \phi^*}\right)^2$$

$$= 2\xi p(\xi)\left[\eta_\delta(\xi)\frac{\partial F^*}{\partial \phi^*}\frac{\partial^2 F^*}{\partial \xi \partial \phi^*} - \frac{d\eta_\delta(\xi)}{d\xi}\left(\frac{\partial F^*}{\partial \phi^*}\right)^2 - \eta_\delta(\xi)\frac{\partial^2 F^*}{\partial \phi^{*2}}\frac{\partial F^*}{\partial \xi}\right] \quad \phi^* > 1$$

with boundary conditions

$$F(\xi, 0) = 0, \quad \frac{\partial F(\xi, 0)}{\partial \phi} = 0, \quad \Theta(\xi, 0) = 1, \quad \Theta(\xi, 1) = 0,$$

$$C_0 F^*(\xi, 1) = F(\xi, 1), \quad C_1 \frac{\partial F^*(\xi, 1)}{\partial \phi^*} = \frac{\partial F(\xi, 1)}{\partial \phi},$$

$$C_2 \frac{\partial^2 F^*(\xi, 1)}{\partial \phi^{*2}} = \frac{\partial^2 F(\xi, 1)}{\partial \phi^2},$$

$$\frac{\partial F^*(0, +\infty)}{\partial \phi^*} = U_e(0)\eta_\delta(0) \quad \text{and} \quad \frac{\partial^2 F^*(\xi, +\infty)}{\partial \phi^{*2}} = 0 \quad \xi > 0.$$

We merge $F(\xi, \phi)$ and $F^*(\xi, \phi^*)$ into a unitary function and introduce a continuation of the definition domain of $\Theta(\xi, \phi)$ to the infinite region as follows

$$g(\xi, \phi) = \begin{cases} F(\xi, \phi) & 0 \leq \phi < 1 \\ F^*(\xi, \phi^*) & \phi = \phi^* > 1 \end{cases}$$

$$\varphi(\xi, \phi) = \begin{cases} \Theta(\xi, \phi) & 0 \leq \phi < 1 \\ 0 & \phi > 1 \end{cases}.$$

It is obvious that $g(\xi, \phi)$ and $\varphi(\xi, \phi)$ are not defined at $\phi = 1$, but there are two limits, left limit and right limit, at $\phi = 1$ for them. In terms of the new variables, the transformed equations in the domain $\xi \geq 0$, $\phi \geq 0$ are

$$\frac{\partial^3 g}{\partial \phi^3} + \alpha(\xi)\eta_\delta(\xi)g\frac{\partial^2 g}{\partial \phi^2} + \beta(\xi)\eta_\delta(\xi)\left[\gamma(\xi)\eta_\delta^2(\xi)H(1 - \phi) - \left(\frac{\partial g}{\partial \phi}\right)^2\right]$$

$$= 2\xi p(\xi)\left[\eta_\delta(\xi)\frac{\partial g}{\partial \phi}\frac{\partial^2 g}{\partial \xi \partial \phi} - \frac{d\eta_\delta(\xi)}{d\xi}\left(\frac{\partial g}{\partial \phi}\right)^2 - \eta_\delta(\xi)\frac{\partial^2 g}{\partial \phi^2}\frac{\partial g}{\partial \xi}\right]$$

Numerical Method for Differentio-Integral Equations$$\frac{1}{P_r}\frac{\partial^2 \varphi}{\partial \phi^2} + \alpha(\xi)\eta_\delta(\xi)g\frac{\partial \varphi}{\partial \phi} = 2\xi p(\xi)\eta_\delta(\xi)\left[\frac{\partial g}{\partial \phi}\frac{\partial \varphi}{\partial \xi} - \frac{\partial \varphi}{\partial \phi}\frac{\partial g}{\partial \xi}\right]$$

with boundary conditions

$$g(\xi, 0) = 0, \quad \frac{\partial g(\xi, 0)}{\partial \phi} = 0, \quad \varphi(\xi, 0) = 1, \quad \varphi(\xi, 1^-) = 0, \quad \varphi(\xi, 1^+) = 0,$$

$$C_0 g(\xi, 1^+) = g(\xi, 1^-), \quad C_1\frac{\partial g(\xi, 1^+)}{\partial \phi} = \frac{\partial g(\xi, 1^-)}{\partial \phi},$$

$$C_2\frac{\partial^2 g(\xi, 1^+)}{\partial \phi^2} = \frac{\partial^2 g(\xi, 1^-)}{\partial \phi^2},$$

$$\frac{\partial g(0, +\infty)}{\partial \phi} = U_e(0)\eta_\delta(0) \quad \text{and} \quad \frac{\partial^2 g(\xi, +\infty)}{\partial \phi^2} = 0 \quad \xi > 0,$$

$$\varphi(\xi, +\infty) = 0$$

where the superscripts − and + mean the left limit and the right limit respectively. Two new conditions for $\varphi(\xi, \phi)$ are added to maintain consistency between the unknown functions and conditions. $H(x)$ is the unit-step function, also known as the Heaviside function or Heaviside's step function, which is usually defined by

$$H(x) = \begin{cases} 0 & x < 0 \\ \frac{1}{2} & x = 0 \\ 1 & x > 0 \end{cases}.$$

Using the same idea, one may also merge the three functions, $F(\xi, \phi)$, $F^*(\xi, \phi^*)$ and $\Theta(\xi, \phi)$, into a unitary function, but it is not helpful in solving the resultant difference equations hereafter because the differential equation for $\Theta(\xi, \phi)$ involves the unknown function $F(\xi, \phi)$.

The above is the main preparatory stage for introducing a merging procedure for the differential equations and conditions. Its principal requirements are based on the five following points:

(i) All unknown functions appear explicitly in both the equations and the conditions.

(ii) Unknown functions, which possess the same physical property and are independent of each other in their differential equations, are merged. These may only include one which is defined over an infinite domain.

(iii) All unknown functions with finite domain - those newly merged and the unmerged remainder - are continued into new unknown functions with infinite domain. In a broad sense, that a function with finite domain is continued into another new function with infinite domain is the same as having the function with finite domain merge with a known solution of an auxiliary equation over an infinite domain.

(iv) The necessary and appropriate conditions to maintain consistency are added.

(v) Boundary conditions at $+\infty$ may be replaced by asymptotic representations or alternatively a sufficiently large finite domain may be chosen, at the outer edge of which, the boundary conditions are understood to be satisfied. For simplicity, the latter approach has been adopted in the present work.

Now the equations are written as a first order system by introducing the new dependent variables $u(\xi,\phi)$, $v(\xi,\phi)$ and $w(\xi,\phi)$ as follows:

$$\frac{\partial g}{\partial \phi} = u$$

$$\frac{\partial u}{\partial \phi} = v$$

$$\frac{\partial v}{\partial \phi} + \alpha(\xi)\eta_\delta(\xi)gv + \beta(\xi)\eta_\delta(\xi)\left[\gamma(\xi)\eta_\delta^2(\xi)H(1-\phi) - u^2\right]$$
$$= 2\xi p(\xi)\left[\eta_\delta(\xi)u\frac{\partial u}{\partial \xi} - \frac{d\eta_\delta(\xi)}{d\xi}u^2 - \eta_\delta(\xi)v\frac{\partial g}{\partial \xi}\right]$$

$$\frac{\partial \varphi}{\partial \phi} = w$$

$$\frac{1}{P_r}\frac{\partial w}{\partial \phi} + \alpha(\xi)\eta_\delta(\xi)gw = 2\xi p(\xi)\eta_\delta(\xi)\left[u\frac{\partial \varphi}{\partial \xi} - w\frac{\partial g}{\partial \xi}\right].$$

Boundary conditions now become

$$g(\xi,0) = 0, \quad u(\xi,0) = 0, \quad \varphi(\xi,0) = 1, \quad \varphi(\xi,1^-) = 0, \quad \varphi(\xi,1^+) = 0,$$

$$C_0 g(\xi,1^+) = g(\xi,1^-), \quad C_1 u(\xi,1^+) = u(\xi,1^-), \quad C_2 v(\xi,1^+) = v(\xi,1^-),$$

$$u(0,\phi_\infty) = U_e(0)\eta_\delta(0) \quad \text{and} \quad v(\xi,\phi_\infty) = 0 \quad \xi > 0, \quad \varphi(\xi,\phi_\infty) = 0.$$

We place an arbitrary rectangular net of points $(\xi_n, \phi_j)$ on $\xi \geq 0$, $0 \leq \phi \leq \phi_\infty$ and use the notation:

$$\xi_0 = 0, \quad \xi_n = \xi_{n-1} + k_n, \quad n = 1, 2, \cdots\cdots;$$
$$\phi_0 = 0, \quad \phi_j = \phi_{j-1} + h_j, \quad j = 1, 2, \cdots, J_1, \cdots, J_2; \quad (2)$$

where $\phi_{J_1} = 1$, $\phi_{J_2} = \phi_\infty$. As a result of the discontinuity for $g(\xi,\phi)$ and non-differentiality of $\varphi(\xi,\phi)$ at $\phi = 1$, the point $\phi = 1$ must be included as a mesh-point. No additional restrictions need be placed on the meshwidths $h_j$ and $k_n$ except this requirement. Because $g(\xi,\phi)$ and $\varphi(\xi,\phi)$ have two values at $\phi = 1$, there are two ways to construct the scheme. One is that the function has two values at the same point $\phi = 1$, that is, the left limit and the right limit. The other is that the point $\phi = 1$ is thought to be two points with zero distance between them, that is, $h_{J_1+1} = 0$, and the function values at the left point and the right point are the left limit and the right limit respectively. Here the former one is preferred and implemented. Note that, for any function $z(\xi,\phi)$, $z_{J_1}^n$ represents the left limit or the right limit hereafter.

If $(g_j^n, u_j^n, v_j^n, \varphi_j^n, w_j^n)$ are to approximate $(g, u, v, \varphi, w)$ at $(\xi_n, \varphi_j)$, the difference approximations are defined, for $1 \leq j \leq J_2$, by

$$\frac{g_j^n - g_{j-1}^n}{h_j} = u_{j-1/2}^n \tag{3}$$

$$\frac{u_j^n - u_{j-1}^n}{h_j} = v_{j-1/2}^n \tag{4}$$

$$\frac{v_j^{n-1/2} - v_{j-1}^{n-1/2}}{h_j} + \alpha_{n-1/2} \left(\eta_\delta(\xi) g v\right)_{j-1/2}^{n-1/2}$$
$$+ \beta_{n-1/2} \gamma_{n-1/2} H(J_1 - j + 1/2) \left(\eta_\delta^3(\xi)\right)_{j-1/2}^{n-1/2} - \beta_{n-1/2} \left(\eta_\delta(\xi) u^2\right)_{j-1/2}^{n-1/2}$$
$$= 2\xi_{n-1/2} p_{n-1/2} \left[ \left(\eta_\delta(\xi) u\right)_{j-1/2}^{n-1/2} \frac{u_{j-1/2}^n - u_{j-1/2}^{n-1}}{k_n} \right.$$
$$\left. - \left(u^2\right)_{j-1/2}^{n-1/2} \frac{\eta_\delta(\xi_n) - \eta_\delta(\xi_{n-1})}{k_n} - \left(\eta_\delta(\xi) v\right)_{j-1/2}^{n-1/2} \frac{g_{j-1/2}^n - g_{j-1/2}^{n-1}}{k_n} \right] \tag{5}$$

$$\frac{\varphi_j^n - \varphi_{j-1}^n}{h_j} = w_{j-1/2}^n \tag{6}$$

$$\frac{1}{P_r} \frac{w_j^{n-1/2} - w_{j-1}^{n-1/2}}{h_j} + \alpha_{n-1/2} \left(\eta_\delta(\xi) g w\right)_{j-1/2}^{n-1/2}$$
$$= 2\xi_{n-1/2} p_{n-1/2} \left[ \left(\eta_\delta(\xi) u\right)_{j-1/2}^{n-1/2} \frac{\varphi_{j-1/2}^n - \varphi_{j-1/2}^{n-1}}{k_n} \right.$$
$$\left. - \left(\eta_\delta(\xi) w\right)_{j-1/2}^{n-1/2} \frac{g_{j-1/2}^n - g_{j-1/2}^{n-1}}{k_n} \right] \tag{7}$$

where $\xi_{n-1/2} = \frac{\xi_n + \xi_{n-1}}{2}$, $\alpha_{n-1/2}$, $\beta_{n-1/2}$, $\gamma_{n-1/2}$ and $p_{n-1/2}$ are the values of $\alpha(\xi)$, $\beta(\xi)$, $\gamma(\xi)$ and $p(\xi)$ at $\xi_{n-1/2}$ respectively. For any function $z(\xi, \phi)$, a notation has been introduced for averages and intermediate values as

$$z_{j-1/2}^n = \frac{z_j^n + z_{j-1}^n}{2}$$

$$z_j^{n-1/2} = \frac{z_j^n + z_j^{n-1}}{2}$$

$$z_{j-1/2}^{n-1/2} = \frac{z_j^n + z_{j-1}^n + z_j^{n-1} + z_{j-1}^{n-1}}{4}.$$

Note that the first, second and fourth equations are centered at $(\xi_n, \phi_{j-1/2})$ while the third and fifth equations are centered at $(\xi_{n-1/2}, \phi_{j-1/2})$. Since the first, second and fourth equations do not have a $\xi$ derivative, they can be differenced about the point $(\xi_n, \phi_{j-1/2})$. It is found in practice that this damps high frequency Fourier error components better than differencing about $(\xi_{n-1/2}, \phi_{j-1/2})$.

The boundary conditions become simply:

$$g_0^n = 0, \quad u_0^n = 0, \quad \varphi_0^n = 1, \quad \varphi_{J_1^-}^n = 0, \quad \varphi_{J_1^+}^n = 0,$$

$$C_0 g_{J_1^+}^n = g_{J_1^-}^n, \quad C_1 u_{J_1^+}^n = u_{J_1^-}^n, \quad C_2 v_{J_1^+}^n = v_{J_1^-}^n, \tag{8}$$

$$u_{J_2}^0 = U_e(0)\eta_\delta(0) \quad \text{and} \quad v_{J_2}^n = 0 \quad n = 1, 2, \cdots, \quad \varphi_{J_2}^n = 0.$$

## 3  Differentio-Integral Reduction Scheme

The same transformation is made as in the merging procedure for differential equations and conditions. The differentio-integral equation is transformed to the following equation

$$H_0 \left[ \left(\frac{\partial \Theta}{\partial \phi}\right)_{\phi=0} + P_r \alpha(\xi)\eta_\delta(\xi) \int_0^1 \frac{\partial F}{\partial \phi} \Theta d\phi \right.$$
$$+ 2\xi P_r p(\xi)\eta_\delta(\xi) \int_0^1 \left( \frac{\partial^2 F}{\partial \phi \partial \xi} \Theta + \frac{\partial F}{\partial \phi} \frac{\partial \Theta}{\partial \xi} \right) d\phi \left. + \alpha(\xi)\eta_\delta(\xi) (F)_{\phi=1} \right]$$
$$+ 2\xi p(\xi)\eta_\delta(\xi) \left( \frac{\partial F}{\partial \xi} \right)_{\phi=1} = 0$$

or

$$H_0 \left[ (w)_{\phi=0} + P_r \alpha(\xi)\eta_\delta(\xi) \int_0^1 u\varphi d\phi + 2\xi P_r p(\xi)\eta_\delta(\xi) \int_0^1 \left( \frac{\partial u}{\partial \xi}\varphi + u\frac{\partial \varphi}{\partial \xi} \right) d\phi \right]$$
$$+ \alpha(\xi)\eta_\delta(\xi) (g)_{\phi=1} + 2\xi p(\xi)\eta_\delta(\xi) \left( \frac{\partial g}{\partial \xi} \right)_{\phi=1} = 0.$$

The main preparatory stage for developing a difference for the differentio-integral equation is a reduction procedure. Its principal requirements are based on the two following points:

(i) The transformations used for the differentio-integral equations are the same as those for the differential equations and conditions.

(ii) The formulation should always ensure that all limits of integration are constants.

The difference approximations are defined by

$$H_0 \left\{ w_0^{n-1/2} + P_r \alpha_{n-1/2} \sum_{j=1}^{J_1} (\eta_\delta(\xi)u\phi)_{j-1/2}^{n-1/2} h_j \right.$$

$$+2\xi_{n-1/2}P_r p_{n-1/2}\sum_{j=1}^{J_1}\left[(\eta_\delta(\xi)\varphi)_{j-1/2}^{n-1/2}\frac{u_{j-1/2}^n - u_{j-1/2}^{n-1}}{k_n}\right.$$

$$\left.+ (\eta_\delta(\xi)u)_{j-1/2}^{n-1/2}\frac{\varphi_{j-1/2}^n - \varphi_{j-1/2}^{n-1}}{k_n}\right]h_j \bigg\} + \alpha_{n-1/2}(\eta_\delta(\xi)g)_{J_1}^{n-1/2}$$

$$+\xi_{n-1/2}p_{n-1/2}\left[\eta_\delta(\xi_n) + \eta_\delta(\xi_{n-1})\right]\frac{g_{J_1}^n - g_{J_1}^{n-1}}{k_n} = 0. \qquad (9)$$

Note that the equation is centered at the line $\xi = \xi_{n-1/2}$.

## 4 Solution of the Difference Equations

The nonlinear difference equations are solved recursively starting with $n = 0$ (on $\xi = \xi_0 = 0$). In the case of $n = 0$, we retain (3), (4) and (6) with $n = 0$ and simply alter (5), (7) and (9) by setting $\xi_{n-1/2} = 0$ and using superscripts $n = 0$ rather than $n-1/2$ in the remaining terms. The resultant difference equations, are then easily solved by the scheme described below and accurate approximations to the solution of the systems of the ordinary differential equations are obtained.

The resultant linear systems of equations can be written in the block arrow-like matrix form as:

$$A^{(i)}\Delta^{(i)} = q^{(i)} \qquad i = 0, 1, 2, \cdots \cdots \qquad (10)$$

where

$$\Delta^{(i)} \equiv \left(\delta_0^{(i)^T}, \cdots, \delta_{J_1^-}^{(i)^T}, \delta_{J_1^+}^{(i)^T}, \cdots, \delta_{J_2}^{(i)^T}, \delta\eta_\delta^{(i)}(\xi_n), \delta S_1, \delta S_2, \delta S_3, \delta S_4\right)^T$$

$$q^{(i)} \equiv \left(0,0,0, r_{1/2}^{(i)^T}, \cdots, r_{J_1-1/2}^{(i)^T}, 0,0,0,0,0, r_{J_1+1/2}^{(i)^T}, \cdots, r_{J_2-1/2}^{(i)^T}, 0, 0, \gamma_\delta^{(i)}, 0, 0, 0, 0\right)^T.$$

The matrix $A^{(i)}$ can be expanded by four extra rows and columns to form a block arrow-like matrix with $5 \times 5$ blocks if we include four extra unknowns $S_1$, $S_2$, $S_3$ and $S_4$ and four extra equations $S_1 = 0$, $S_2 = 0$, $S_3 = 0$ and $S_4 = 0$. The ordering of equations is (i) the three boundary conditions (8) at $\phi = 0$, (ii) the equations (3)-(7) at the centered location $j = 1/2, \cdots, J_1 - 1/2$, (iii) the five boundary conditions (8) at $\phi = 1$, (iv) the equations (3)-(7) at the centered location $j = J_1+1/2, \cdots, J_2-1/2$, (v) the two equations (8) at $\phi = \phi_\infty$, (vi) the equation (9) and (vii) the four dummy variable equations. We now have $5J_2+15$ equations and unknowns and the matrix $A^{(i)}$ has the form

$$A^{(i)} = \begin{pmatrix} A_0 & C_0 & & & & & & & & D_0 \\ B_1 & A_1 & C_1 & & & & & & & D_1 \\ & B_2 & A_2 & C_2 & & & & & & D_2 \\ & & \cdot & \cdot & \cdot & & & & & \cdot \\ & & & \cdot & \cdot & \cdot & & & & \cdot \\ & & & & \cdot & \cdot & \cdot & & & \cdot \\ & & & & & B_{N-1} & A_{N-1} & C_{N-1} & & D_{N-1} \\ & & & & & & B_N & A_N & & D_N \\ E_0 & E_1 & E_2 & \cdot & \cdot & \cdot & E_{N-1} & E_N & & A_{N+1} \end{pmatrix}$$

where $N = J_2 + 1$. The matrix $A^{(i)}$ is called a block arrow-like matrix here due to its shape. A suitable algorithm for solving the matrix equation in (10) is the direct factorization method of block arrow-like matrix [6-12]. Algorithmically this is simply a modification of the usual solution of a block tridiagonal system to include the block arrow-like matrix and can be made efficient by taking account of the zeros appearing in matrices, comparing with the classical methods [13-16].

## 5 An Example of Application

We choose

$$\alpha(\xi) = \frac{1 + 24\xi}{(1+\xi)^{1/4}(1+16\xi)^{3/4}}, \quad \beta(\xi) = \frac{\xi(17 + 32\xi)}{2(1+\xi)^{5/4}(1+16\xi)^{3/4}},$$

$$\gamma(\xi) = \frac{4(1+\xi)^{1/2}(1+16\xi)^{1/2}}{17 + 32\xi}, \quad p(\xi) = \frac{(1+16\xi)^{1/4}}{(1+\xi)^{1/4}},$$

$U_e(0) = 1$ and the constants $C_0 = \frac{1}{\lambda\omega}$, $C_1 = \frac{1}{\lambda^2}$, $C_2 = \frac{1}{\lambda^3\omega}$ where $\lambda$ and $\omega$ are two physical constants. Integrating (1) gives

$$U_e(\xi) = \frac{1}{(1+\xi)^{1/4}(1+16\xi)^{1/4}}.$$

The choice yields the equations for combined gravity body force and forced convection in laminar film condensation [5]. Consequently, the present problem in fact involves four physical constants $H_0$, $P_r$, $\lambda$ and $\omega$. Before the numerical solutions of the foregoing equations are discussed, it is worthwhile to consider the extreme case when the liquid film is very thin. In this case, perturbation techniques may be used and the following approximate formulas are obtained in the form

$$H_0 = \frac{0.2348}{\lambda^3\omega}\eta_\delta^3(0) + O\left(\eta_\delta^5(0) + \xi\right) \quad \text{as} \quad \xi \longrightarrow 0$$
$$H_0 = \eta_\delta^4(+\infty) + O\left(\eta_\delta^8(+\infty) + \xi^{-1/2}\right) \quad \text{as} \quad \xi \longrightarrow +\infty. \quad (11)$$

The detailed mathematical analysis and physical background in this problem have been explained elsewhere [5]. When $H_0 = 0.008191$, $P_r = 10$, $\lambda = 1$ and $\omega = 10$ are chosen, it is found that the liquid film is very thin. The results can be compared with the analytical approximate formulas (11), so it is employed as a test case. A coarse grid of dimensions $42 \times 80$ is taken on the $(\xi, \phi)$ domain with each cell being divided into 1, 2, 3, 4 sub-cells for the two directions simultaneously and the iteration error is assigned as $0.5 \times 10^{-13}$. The values of $\xi_n$ used are shown in Table 1. Note that the point $\phi = 1$, which is equivalent to the interface between liquid phase and vapor phase, must be included as a coarse mesh-point.

The most sensitive and the most important quantity in such calculations is $\eta_\delta(\xi)$ from which the liquid film thickness of condensate layer is determined.

Numerical Method for Differentio-Integral Equations

**Table 1.**

| $\xi$ | $\eta_\delta(\xi)$ | $\xi$ | $\eta_\delta(\xi)$ | $\xi$ | $\eta_\delta(\xi)$ |
|---|---|---|---|---|---|
| 0.0 | 0.661 | 0.3 | 0.414 | 15 | 0.305 |
| $10^{-6}$ | 0.661 | 0.4 | 0.396 | 20 | 0.303 |
| $10^{-5}$ | 0.661 | 0.6 | 0.374 | 30 | 0.302 |
| $10^{-4}$ | 0.661 | 0.8 | 0.360 | 60 | 0.301 |
| $5 \times 10^{-4}$ | 0.659 | 1.0 | 0.351 | 100 | 0.301 |
| $10^{-3}$ | 0.657 | 1.3 | 0.341 | 250 | 0.301 |
| $4 \times 10^{-3}$ | 0.645 | 1.6 | 0.335 | $10^3$ | 0.301 |
| 0.01 | 0.624 | 2.0 | 0.329 | $10^4$ | 0.301 |
| 0.025 | 0.585 | 2.5 | 0.324 | $10^5$ | 0.301 |
| 0.05 | 0.542 | 3.2 | 0.319 | $10^6$ | 0.301 |
| 0.075 | 0.513 | 4.0 | 0.316 | $10^8$ | 0.301 |
| 0.1 | 0.492 | 5.0 | 0.312 | $10^{10}$ | 0.301 |
| 0.15 | 0.462 | 7.0 | 0.309 | $10^{16}$ | 0.301 |
| 0.2 | 0.441 | 10 | 0.306 | $10^{24}$ | 0.301 |

Thus only this quantity shall be examined in the discussion of results. All other computed quantities behaved similarly. Table 1 depicts the flavor of these results. From the convergence of the extrapolation process described in the last section, the absolute error is $2.8 \times 10^{-8}$. From this data, the results seem accurate to 7 decimal places which is more than sufficient for most purposes.

The approximate formulas (11) predict that $\eta_\delta(0) \approx 0.704$ and $\eta_\delta(+\infty) \approx 0.301$ for $H_0 = 0.008191$, $P_r = 10$, $\lambda = 1$ and $\omega = 10$, whereas the data from Table 1 display that $\eta_\delta(0) = 0.661$ and $\eta_\delta(+\infty) = 0.301$. Despite the fact which higher-order terms have been omitted when $\eta_\delta$ in (11) is calculated, a good agreement is obtained between the purely numerical computations and the analytical approximate formulas for thin film theory, even though $\eta_\delta(0) = 0.661$ is a moderate, rather than a thin value.

## 6 Conclusion

We have demonstrated how to obtain the results of high accuracy for systems of differentio-integral equations having multilayer flows by modifying the Keller Box method and incorporating extrapolation. The technique can be used as the basis for more complex problems.